\documentclass[%
 reprint,
 amsmath,amssymb,
 aps,
 prl,
]{revtex4-2}

\usepackage{graphicx}
\usepackage{dcolumn}
\usepackage{bm}

\usepackage{color}

\begin{document}

\preprint{APS/123-QED}

\title{
Angular dependence of magnetoresistance and planar Hall effect in semimetals in strong magnetic fields}

\author{Akiyoshi Yamada}
\author{Yuki Fuseya}%
\affiliation{%
 Department of Engineering Science, University of Electro-Communications, Chofu, Tokyo 182-8585, Japan\\
}%




\date{\today}

\begin{abstract}
The semiclassical transport theory is especially powerful for investigating galvanomagnetic effects. Generally, the semiclassical theory is applicable only in weak fields because it does not consider Landau quantization. Herein, we extend the conventional semiclassical theory by considering Landau quantization through the field dependence of carrier density in semimetals. The extended semiclassical theory is applicable even in strong fields, where Landau quantization is noticeable.
Using this new approach, we explain the qualitative change in the angular dependence of transverse magnetoresistance (TMR), anisotropic magnetoresistance (AMR), and planar Hall effect (PHE) in bismuth with an increase in the magnetic field. This unveils the puzzle of nontrivial field-induced changes in TMR, AMR, and PHE observed recently in semimetal bismuth. 

\end{abstract}
\maketitle


The galvanomagnetic effect, in which the direction and amplitude of current are changed by a magnetic field, is one of the oldest and most fundamental subjects in solid-state physics. Every textbook of solid-state physics provides its introductory explanation, and some books discuss it in detail \cite{Beer1963,Pippard_book}. Despite it being discovered centuries ago, the galvanomagnetic effect continues to drive novel physics even today. Recently, the galvanomagnetic effect has attracted significant attention in connection with  topological condensed-matter physics \cite{Nielsen1983,Armitage2018}. For example, the possibilities of observing the chiral anomaly in solids are being actively discussed, such as in negative magnetoresistance \cite{Nielsen1983,Son2013,Huang2015,Li2018} and in planar Hall effect (PHE) \cite{Burkov2017,Nandy2017,Li2018,Wu2018,Liang2018,Chen2018,PLi2018,Singha2018,Kumar2018,Yang2020}.

The semiclassical transport theory has proved to be very useful in explaining the galvanomagnetic effect \cite{Beer1963,Owada2018,Armitage2018}.
For example, it could successfully explain the complex galvanomagnetic effect in bismuth, a multivalley semimetal with Dirac electrons \cite{Dresselhaus1971,Edelman1976,Issi1979,Fuseya2015,Zhu2018}. The angular dependences of transverse magnetoresistance (TMR), anisotropic magnetoresistance (AMR), and PHE in Bi have been accurately explained by the semiclassical theory for weak fields ($\lesssim 1$ T) \cite{Collaudin2015,Zhu2018,Yang2020}. 
However, the angular dependence exhibits an essential transformation by the magnetic field from weak to strong. (The experimental details are provided later in the manuscript.)
The semiclassical theory has failed to explain these field-induced transformations in the angular dependences \cite{Collaudin2015,Zhu2017,Yang2020}. 
This leads to a simple question. Does the apparent failure of the semiclassical theory indicate the appearance of novel physics, or implies the limitation of the semiclassical theory?

Regarding the possibility of novel physics, Yang {\it et al.} argued the possibility that the field-induced transformation in Bi originates from the effect of the chiral anomaly in Weyl and Dirac electrons \cite{Yang2020}. However, the same study also pointed out that a phase shift of $\pi/2$ in the bisectrix channel of PHE did not agree with the chiral anomaly model.
Typically, the semiclassical theory is not applicable in strong fields, because it does not consider Landau quantization. If this difficulty is overcome, it can reveal the origin of the field-induced transformation in the galvanomagnetic effect in Bi.

\begin{figure}[tb]
\includegraphics[width=7cm]{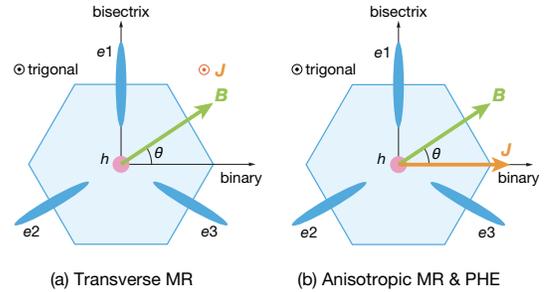}
\caption{\label{fig0} Configuration of the current $\bm{J}$ and magnetic field $\bm{B}$ for (a) TMR, and for (b) AMR and PHE.
One circle and three ellipsoids express the hole and the electron pockets of the Fermi surface in Bi, respectively.}
\end{figure}

This study aims to develop a new theoretical technique to calculate the galvanomagnetic effect in the strong magnetic field, and to clarify the origin of the nontrivial field-induced transformation in the angular dependence of TMR, AMR, and PHE in Bi.
We extend the semiclassical transport theory by considering Landau quantization through the field dependence of carrier density, which is crucial for semimetals in strong fields. Using this approach, we can expand the scope of application of the semiclassical theory to the strong field region. The results of our modified theoretical approach exhibit a field-induced transformation in the angular dependence of TMR, AMR, and PHE, all of which agree well with the experiments. 

The electronic structure and Landau quantization of Bi are accurately known owing to a number of previous studies \cite{Dresselhaus1971,Fuseya2015,Zhu2018}. There is no ambiguity in the electric structure. Therefore, Bi provides an ideal environment to examine the new theory.
After verifying the successful implementation of our theory to explain the galvanomagnetic effect in Bi in strong fields, we can apply it to various topological materials. 

Before we discuss the theory, let us summarize the experimental results on the angular dependence of TMR, AMR, and PHE in Bi. 
TMR in Bi exhibits a large angular dependence, where the magnetic field is rotated in the binary-bisectrix plane with the current along the \emph{trigonal} axis [Fig. \ref{fig0} (a)]. The angular dependence is sufficiently large to be observed at 300 K \cite{Zhu2012b}. This highly anisotropic TMR originates from the extremely high mobility of Bi ($\mu \sim 10^8$ cm$^2$/Vs) \cite{Hartman1969}. The TMR is anisotropic only in weak fields ($\lesssim 1$ T) and the anisotropy disappears in strong fields ($\gtrsim 10$ T), i.e., the TMR becomes nearly isotropic \cite{Collaudin2015,Zhu2017}. This is rather surprising because the anisotropy should be enhanced by the magnetic field.
(Note that the loss of three-fold symmetry in TMR has been observed in strong fields \cite{Zhu2012b,Collaudin2015}; however, here we concentrate on the behavior of a large component that holds the three-fold symmetry.)
Remarkable amplitudes of AMR and PHE are observed with the current along the \emph{binary} axis by rotating the magnetic field in the same plane as that of TMR [Fig. \ref{fig0} (b)] \cite{Liang2018,Yang2020}. 
In weak fields ($\lesssim 1$ T), both AMR and PHE have two components of angular oscillation with period $\pi/2$ and $\pi$. In strong fields ($\gtrsim 10$ T), however, the angular oscillation with period $\pi /2$ disappears. We primarily address these nontrivial field-induced phenomena in this study.

Each part of the theory---the Landau levels and carrier density of Bi \cite{Smith1964,Hiruma1983,Zhu2011}, and the semiclassical theory for multivalley systems \cite{Aubrey1971,Roth1992,Askerov1994}---has been already well established in previous studies. The novelty of this work lies in combining these parts to present a unified theory for examining TMR, AMR, and PHE simultaneously.

TMR, AMR and PHE are all calculated from the resistivity tensor $\hat{\rho}$, which is given by the inverse of the conductivity tensor $\hat\rho={\hat\sigma}^{-1}$. For multivalley systems, the total conductivity tensor is given by the summation of each valley conductivity $\hat{\sigma} =\sum_i\hat{\sigma}^{(i)}$ \cite{Aubrey1971}.
Each conductivity tensor is written as 
\begin{equation}
\label{cond}
\hat{\sigma}^{(i)} = en_i\left(\hat{\mu}_i^{-1}\pm {\hat{B}}\right)^{-1}, 
\end{equation}
where $e>0$ is the elementary charge and $\hat{B}$ is the magnetic field tensor \cite{Mackey1969,Aubrey1971}. $n_i$ and $\hat{\mu}_i$ are the carrier density and the mobility tensor of each valley, respectively. The sign $\pm$ corresponds to the sign of charge.
It is naively expected that Eq. \eqref{cond} is not applicable in the strong field region, because it was obtained without Landau quantization.
However, it has been shown that the results from an extended semiclassical approach agree significantly well with the fully quantum approach by the Kubo formula except for the quantum oscillation \cite{Owada2018}. Here, the extended semiclassical approach considers Landau quantization through the field dependence of carrier density. This approach has not yet been applied to specific materials. In the present work, we employ the extended semiclassical theory with the Landau levels and the field- and angular-dependences of carrier density in Bi, which have been accurately determined by several studies \cite{Smith1964,Hiruma1983,Zhu2011,Zhu2018}.
The field-induced angular dependence of carrier density plays a crucial role in solving the anomalies in strong fields.

One hole pocket is located at the $T$ point along the trigonal axis, and three electron pockets are located at three equivalent $L$ points (Fig. \ref{fig0}). 
The electrons around the $L$ point can be well described as the Dirac electrons with an additional $g$-factor, which originates from the multiband effect of spin-orbit coupling \cite{Wolff1964,Baraff1965,Fuseya2015b,Izaki2019}. Its Landau quantized energy is given as \cite{Zhu2011}
\begin{align}
\label{Dirac}
\epsilon_{n,\sigma}(k_z)&= \sqrt{\Delta^2 +2\Delta \xi_{n, \sigma}(k_z ) } 
		+ \frac{\sigma g'\mu_B B}{2},
		\nonumber\\
		\xi_{n, \sigma}(k_z)&=\left(n+\frac{1}{2} +\frac{\sigma}{2}\right)\hbar\omega_c + \frac{\hbar^2 k^2_z}{2m_z},
\end{align}
where $n$ is the Landau level index, and $\sigma =\pm$ corresponds to the degree of freedom of the Kramers doublet. $\Delta$ is the half of the band-gap and $\omega_c$ is the cyclotron frequency. $g'$ is the additional $g$-factor, and $\mu_B$ is the Bohr magneton. $k_z$ and $m_z$ are the wavenumber and effective mass along the magnetic field direction, respectively.
The holes around the $T$ point can be well approximated as a nearly free particle with a modified $g$-factor \cite{Smith1964,Fuseya2015b} (for details see \cite{SeeSM}). 
The Fermi energy is determined in order to satisfy the charge neutrality condition:
$n_h(B) =\sum_{i=1{\text -}3} n_{ei}(B)$, where $n_h(B)$ and $n_{ei}(B)$ are the carrier densities of holes and electrons, respectively.
Finally, we calculate the magneto-conductivity tensors by substituting $n_i$ in Eq. \eqref{cond} with the computed field-dependent carrier density. This theoretical approach enables the calculation of the galvanomagnetic effect even at strong fields, where Landau quantization is noticeable \cite{Owada2018}.

\begin{figure}[tb]
\includegraphics{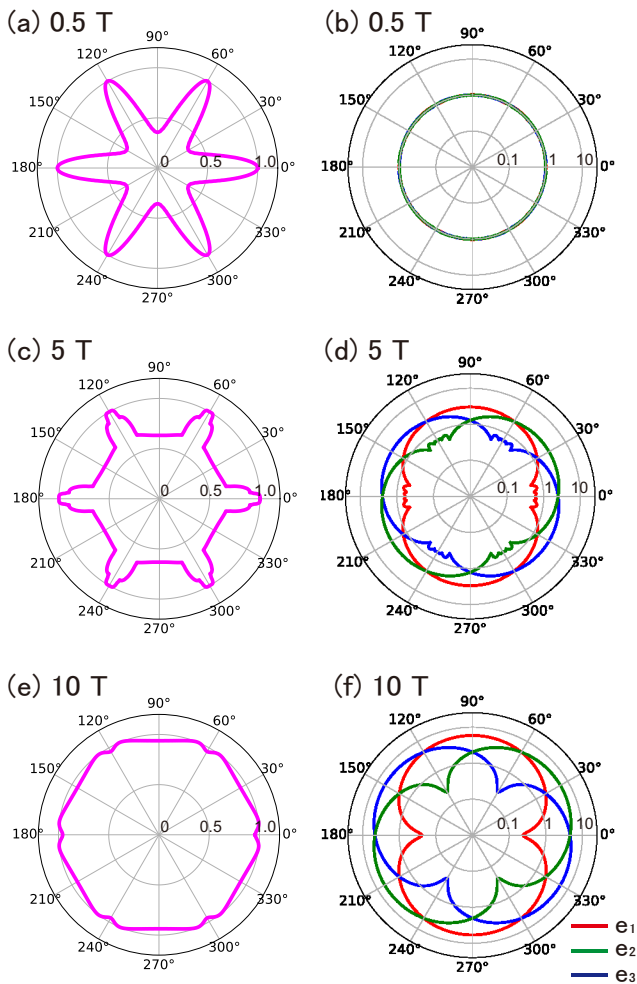}
\caption{\label{fig1} (Left) Angle dependence of inverse TMR with the current along the trigonal axis, $\rho_{33}^{-1}(\bm{B})$, at 0.5, 5, and 10 T. 
For easy comparison with the experiment \cite{Collaudin2015}, the values of $\rho_{33}^{-1}$ are normalized by its maximum values.
(Right) Angle dependence of electron carrier densities, $n_{\rm e1, e2, e3} (\bm{B})$, which are normalized by those in the weak-field limit , $n_{ei}(0.1 {\rm T})=9.7\times 10^{16}$ cm$^{-3}$.
$0^\circ$ and $90^\circ$ correspond to the binary and bisectrix axis, respectively.}
\end{figure}

The left panels of Fig. \ref{fig1} are the polar plots of the inverse of TMR $\rho_{33}^{-1} (\theta)$, where the field is rotated in the binary-bisectrix plane with the current along the trigonal axis [Fig. \ref{fig0} (a)]. Subscripts 1, 2, and 3 correspond to binary, bisectrix, and trigonal directions, respectively. (Here, we plot the inverse of TMR to make the comparison with the experiment easier \cite{Collaudin2015}. This plot makes the comparison with the theoretical carrier density easier as well.) We used the electron mobilities: $\mu_1=11000$, $\mu_2=300$, $\mu_3=6700$, $\mu_4=-710$, and the hole mobilities: $\nu_1=2200$, and $\nu_3=350$ (in the units of T$^{-1}=10^4$ cm$^2$/V s), which were obtained by Hartman for bulk Bi at 4.23 K \cite{Hartman1969}.
In weak fields, $\rho_{33}^{-1}(\theta)$ takes the maxima (minima) for $B\parallel$ binary (bisectrix). The highly anisotropic $\rho_{33}^{-1}(\theta)$ in weak fields originates from the anisotropy of mobilities. In particular, $\rho_{33}^{-1}$ is proportional to the effective mass \emph{perpendicular} to the magnetic field, because the conductivity of the electron pocket along the bisectrix axis ($e1$ in Fig. \ref{fig0}) is given as $\sigma_{33}^{e1}\simeq e n_{e1}/\left(\mu_2 \cos^2 \theta + \mu_1 \sin^2 \theta \right)B^2$, and the mobilities are inversely proportional to the effective mass. 
As the field increases, the star-shaped peaks at the binary become less prominent and vanish at 10 T [Fig. \ref{fig1} (e)]. Here, the angular dependence of TMR is almost isotropic. 
This regression of anisotropy is perfectly consistent with the experiment, especially at low temperatures \cite{Collaudin2015,Zhu2017}.

The qualitative change in the anisotropy can be easily understood by considering the angular dependence of carrier density of electrons, $n_{ei} (\theta)$, shown in the right panels in Fig. \ref{fig1}. 
In weak fields, $n_{ei}(\theta)$ is isotropic, so that the anisotropy of TMR originates solely from the anisotropy of mobilities ($\mu_1 \gg \mu_2$). In strong fields, on the other hand, $n_{ei}(\theta)$ becomes anisotropic.
$n_{ei}(\theta)$ takes its maximum (minimum) values for $B\parallel $ bisectrix (binary). The anisotropy of $n_{ei} (\theta)$ is orthogonal to that of $\rho_{33}^{-1}(\theta)$, where they compensate each other. This is the reason why the TMR becomes isotropic at high fields.

\begin{figure}[tb]
\includegraphics[width=7cm]{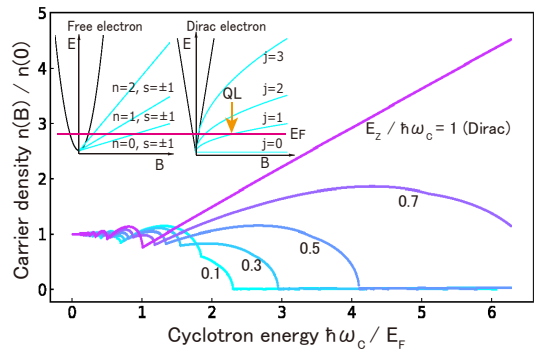}
\caption{\label{fig2} Magnetic field dependence of carrier density for different Zeeman energies, which characterizes the amplitude of the spin-orbit coupling. The insets show the field dependence of the Landau levels for the free and Dirac electrons.}
\end{figure}

The origin of the anisotropy of $n_{ei}(\theta)$ is related to (i) the characteristics of Dirac electrons in semimetals, and (ii) the anisotropy of effective mass.
Let us discuss the first possible origin. In semimetals, where electrons and holes coexist even at zero temperature, $n_{ei}$ can change to a great extent as long as it maintains charge neutrality \cite{Smith1964,Zhu2011,Zhu2018}.
As shown in Fig. \ref{fig2}, $n_{ei}$ begins to acquire a charge when the field reaches the quantum limit ($\hbar \omega_c \sim E_F$), where all electrons are confined to the lowest Landau level. (The details of the calculation are given in Ref. \cite{SeeSM}.) In the case of Dirac electrons ($E_z/\hbar \omega_c=1$), the lowest Landau level is barely affected by the magnetic field (cf. the inset in Fig. \ref{fig2}). Thus, $n_{ei}$ increases linearly in $B$ due to Landau degeneracy. On the other hand, in the case of nearly free electrons ($E_z/\hbar \omega_c \ll1$), $n_{ei}$ decreases with the field, because the energy of the lowest Landau level increases with the magnetic field (the inset in Fig. \ref{fig2}). Therefore, the large increase in the carrier density is a characteristic of Dirac electrons in semimetals with large spin-orbit coupling.
Now, we discuss the second probable origin. The increase in $n_{ei}$ depends on the inverse of cyclotron mass $m_c$, because the field value in the quantum limit is roughly given by $E_F \simeq \hbar \omega_c \propto 1/m_c$. Here, $m_c$ is given by the effective mass \emph{perpendicular} to the field.
The anisotropy of $\rho_{33}^{-1} (\theta) $ in weak fields, which is proportional to $m_c$, is canceled out by the anisotropy of $n_{ei}$ in strong fields. Consequently, the isotropic TMR in strong fields is the inherent characteristic of semimetals with Dirac electrons.


\begin{figure}[tb]
\includegraphics{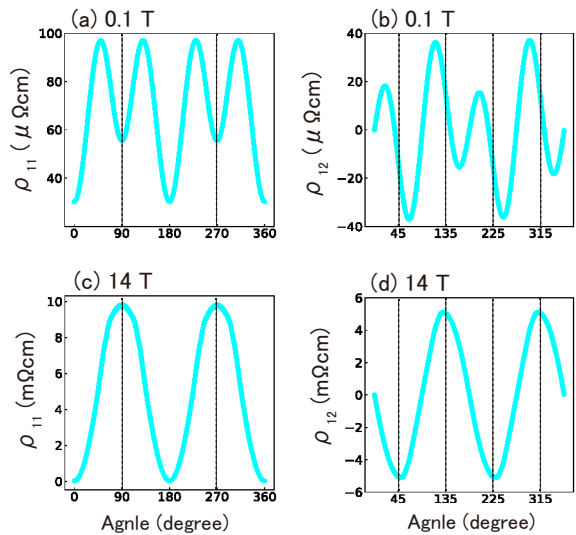}
\caption{\label{fig3} Angular dependence of $\rho_{11}$ (AMR) and $\rho_{12}$ (PHE) with the current along the binary axis, assuming the mobilities for a thin film.}
\end{figure}

We calculate $\rho_{11}$ (AMR) and $\rho_{12}$ (PHE) by the same extended semiclassical theory. The arrangement of $\bm{J}$ and $\bm{B}$ is shown in Fig. \ref{fig0} (b). We used the mobility for a thin film obtained by Yang {\it et al.}: $\mu_1=142.8$, $\mu_2=1.99$, $\mu_3=32.7$, $\mu_4=-3.38$, $\nu_1=18.8$, and $\nu_3=1.57$ (in T$^{-1}$) \cite{Yang2020}. In Fig. \ref{fig3}, we assumed that the charge neutrality is violated as $[n(B)-p(B)]/n(0)=0.265$ according to the experimental report \cite{Yang2020}. Note that the following results are essentially unchanged even if we change the degrees of the violation \cite{SeeSM}.
In weak fields [Figs. \ref{fig3} (a), (b)], both $\rho_{11} (\theta)$ and $\rho_{12}(\theta)$ show four maxima, i.e., the angular oscillation has two components with periods $\pi$ and $\pi/2$. These properties have been already reported by Yang {\it et al.} based on the conventional semiclassical theory \cite{Yang2020}. They also pointed out that the experimental results in strong fields---the peak of period $\pi/2$ disappears---are impossible to be fitted by the conventional semiclassical theory. That is why they argued the possible scenario of chiral anomaly.
In contrast, our results for strong fields [Fig. \ref{fig3} (c), (d)] show that the angular oscillation of the period $\pi/2$ disappears both in $\rho_{11} (\theta)$ and $\rho_{12} (\theta)$, which can explain the experimentally observed field-induced transformation. It is clear that the field- and angular-dependence of $n_{ei}$ play a crucial role for the field-induced transformation.
Even in AMR and PHE, the angular-dependence of $n_{ei}(\theta)$ in strong fields is the same as that in TMR (Fig \ref{fig0}, \ref{fig1}). Therefore, the field-induced transformation in AMR and PHE shares the same origin as in TMR, which have never been previously pointed out.


\begin{figure}[tb]
\includegraphics{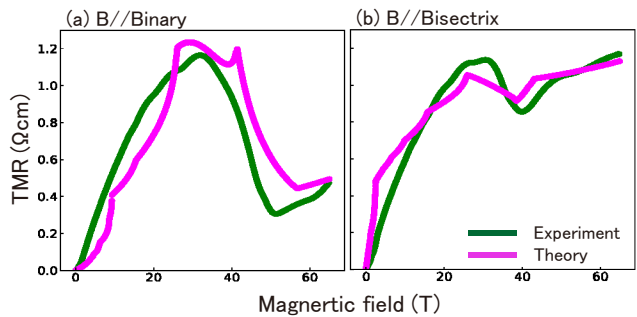}
\caption{\label{fig4} TMR as a function of magnetic field along the (a) binary and (b) bisectrix axis. The theoretical results are obtained by the extended semiclassical theory with the field dependence of mobility. The experimental data are from Ref. \cite{Zhu2017}.}
\end{figure}
We revealed that the nontrivial field-induced transformations of TMR, AMR, and PHE are all well explained by our extended semiclassical theory in a unified manner. However, this approach is not sufficient to obtain quantitative agreements with the measured amplitude of galvanomagnetic effects. For example, the theoretical value of TMR was estimated to be two orders larger than the experimental value at 10 T even when we considered the field-dependent carrier density. 
This discrepancy can be corrected by considering the field dependence of mobilities.
We hypothesized several forms of the field-dependent mobility and found that the functional form of $\hat{\mu} = \hat{\mu}_0/(1+\gamma_i B)$ can well fit the TMR from weak fields to strong fields, as shown in Fig. \ref{fig4}. (The details are provided in \cite{SeeSM}.)
The theoretical results of TMR both for $B \parallel$ binary and bisectrix axis agree well quantitatively with those of experiments including the sudden drop at around 40 T, which is due to the evaporation of Dirac electrons \cite{Zhu2017}.
Although the microscopic derivation of this functional form is still missing, such field dependence may be derived by the guiding center diffusion \cite{Song2015} or quantum correction to the relaxation time \cite{Abrikosov1969,Mahan1983}. 

In summary, there are three aspects to our conclusions. 
First, we proposed a method to extend the scope of application of the semiclassical theory for explaining galvanomagnetic effects up to strong fields. The key components of the modified theory are Landau quantization and the field dependence of carrier density.
Second, we showed the crucial role of the angular dependence of carrier density in strong fields. This property originates from the characteristics of semimetals with Dirac electrons.
Third, we pointed out that the nontrivial field-induced transformation observed in the TMR, AMR, and PHE in Bi can be naturally explained by the angular dependence of carrier density in strong fields. We showed that nontrivial behavior does not indicate the appearance of novel topological physics, such as the chiral anomaly.
We have established these three aspects by employing our theory for Bi.
These three aspects are not only valid for Bi, but can also be applied to various semimetals. 
Using this approach, we can further improve the accuracy of the analysis of galvanomagnetic effects and identify new phenomena, especially in topological semimetals.

\begin{acknowledgments}
We thank K. Behnia, B. Fauqu\'e, and Z. Zhu for providing their experimental data and for useful discussions.
This work is supported by JSPS KAKENHI Grants No. 19H01850 and 18KK0132.
\end{acknowledgments}

\bibliography{MR}

\begin{thebibliography}{41}%
\makeatletter
\providecommand \@ifxundefined [1]{%
 \@ifx{#1\undefined}
}%
\providecommand \@ifnum [1]{%
 \ifnum #1\expandafter \@firstoftwo
 \else \expandafter \@secondoftwo
 \fi
}%
\providecommand \@ifx [1]{%
 \ifx #1\expandafter \@firstoftwo
 \else \expandafter \@secondoftwo
 \fi
}%
\providecommand \natexlab [1]{#1}%
\providecommand \enquote  [1]{``#1''}%
\providecommand \bibnamefont  [1]{#1}%
\providecommand \bibfnamefont [1]{#1}%
\providecommand \citenamefont [1]{#1}%
\providecommand \href@noop [0]{\@secondoftwo}%
\providecommand \href [0]{\begingroup \@sanitize@url \@href}%
\providecommand \@href[1]{\@@startlink{#1}\@@href}%
\providecommand \@@href[1]{\endgroup#1\@@endlink}%
\providecommand \@sanitize@url [0]{\catcode `\\12\catcode `\$12\catcode
  `\&12\catcode `\#12\catcode `\^12\catcode `\_12\catcode `\%12\relax}%
\providecommand \@@startlink[1]{}%
\providecommand \@@endlink[0]{}%
\providecommand \url  [0]{\begingroup\@sanitize@url \@url }%
\providecommand \@url [1]{\endgroup\@href {#1}{\urlprefix }}%
\providecommand \urlprefix  [0]{URL }%
\providecommand \Eprint [0]{\href }%
\providecommand \doibase [0]{https://doi.org/}%
\providecommand \selectlanguage [0]{\@gobble}%
\providecommand \bibinfo  [0]{\@secondoftwo}%
\providecommand \bibfield  [0]{\@secondoftwo}%
\providecommand \translation [1]{[#1]}%
\providecommand \BibitemOpen [0]{}%
\providecommand \bibitemStop [0]{}%
\providecommand \bibitemNoStop [0]{.\EOS\space}%
\providecommand \EOS [0]{\spacefactor3000\relax}%
\providecommand \BibitemShut  [1]{\csname bibitem#1\endcsname}%
\let\auto@bib@innerbib\@empty
\bibitem [{\citenamefont {Beer}(1963)}]{Beer1963}%
  \BibitemOpen
  \bibfield  {author} {\bibinfo {author} {\bibfnamefont {A.}~\bibnamefont
  {Beer}},\ }\href {https://books.google.co.jp/books?id=Q68PAQAAMAAJ} {\emph
  {\bibinfo {title} {Galvanomagnetic effects in semiconductors}}},\ Solid State
  Physics Series\ (\bibinfo  {publisher} {Academic Press},\ \bibinfo {year}
  {1963})\BibitemShut {NoStop}%
\bibitem [{\citenamefont {Pippard}(1989)}]{Pippard_book}%
  \BibitemOpen
  \bibfield  {author} {\bibinfo {author} {\bibfnamefont {A.}~\bibnamefont
  {Pippard}},\ }\href {https://books.google.co.jp/books?id=D5XHMARd2ocC} {\emph
  {\bibinfo {title} {Magnetoresistance in Metals}}},\ Cambridge Studies in Low
  Temperature Physics\ (\bibinfo  {publisher} {Cambridge University Press},\
  \bibinfo {year} {1989})\BibitemShut {NoStop}%
\bibitem [{\citenamefont {Nielsen}\ and\ \citenamefont
  {Ninomiya}(1983)}]{Nielsen1983}%
  \BibitemOpen
  \bibfield  {author} {\bibinfo {author} {\bibfnamefont {H.}~\bibnamefont
  {Nielsen}}\ and\ \bibinfo {author} {\bibfnamefont {M.}~\bibnamefont
  {Ninomiya}},\ }\bibfield  {title} {\bibinfo {title} {The adler-bell-jackiw
  anomaly and weyl fermions in a crystal},\ }\href
  {https://doi.org/https://doi.org/10.1016/0370-2693(83)91529-0} {\bibfield
  {journal} {\bibinfo  {journal} {Physics Letters B}\ }\textbf {\bibinfo
  {volume} {130}},\ \bibinfo {pages} {389 } (\bibinfo {year}
  {1983})}\BibitemShut {NoStop}%
\bibitem [{\citenamefont {Armitage}\ \emph {et~al.}(2018)\citenamefont
  {Armitage}, \citenamefont {Mele},\ and\ \citenamefont
  {Vishwanath}}]{Armitage2018}%
  \BibitemOpen
  \bibfield  {author} {\bibinfo {author} {\bibfnamefont {N.~P.}\ \bibnamefont
  {Armitage}}, \bibinfo {author} {\bibfnamefont {E.~J.}\ \bibnamefont {Mele}},\
  and\ \bibinfo {author} {\bibfnamefont {A.}~\bibnamefont {Vishwanath}},\
  }\bibfield  {title} {\bibinfo {title} {Weyl and dirac semimetals in
  three-dimensional solids},\ }\href
  {https://doi.org/10.1103/RevModPhys.90.015001} {\bibfield  {journal}
  {\bibinfo  {journal} {Rev. Mod. Phys.}\ }\textbf {\bibinfo {volume} {90}},\
  \bibinfo {pages} {015001} (\bibinfo {year} {2018})}\BibitemShut {NoStop}%
\bibitem [{\citenamefont {Son}\ and\ \citenamefont {Spivak}(2013)}]{Son2013}%
  \BibitemOpen
  \bibfield  {author} {\bibinfo {author} {\bibfnamefont {D.~T.}\ \bibnamefont
  {Son}}\ and\ \bibinfo {author} {\bibfnamefont {B.~Z.}\ \bibnamefont
  {Spivak}},\ }\bibfield  {title} {\bibinfo {title} {Chiral anomaly and
  classical negative magnetoresistance of weyl metals},\ }\href
  {https://doi.org/10.1103/PhysRevB.88.104412} {\bibfield  {journal} {\bibinfo
  {journal} {Phys. Rev. B}\ }\textbf {\bibinfo {volume} {88}},\ \bibinfo
  {pages} {104412} (\bibinfo {year} {2013})}\BibitemShut {NoStop}%
\bibitem [{\citenamefont {Huang}\ \emph {et~al.}(2015)\citenamefont {Huang},
  \citenamefont {Zhao}, \citenamefont {Long}, \citenamefont {Wang},
  \citenamefont {Chen}, \citenamefont {Yang}, \citenamefont {Liang},
  \citenamefont {Xue}, \citenamefont {Weng}, \citenamefont {Fang},
  \citenamefont {Dai},\ and\ \citenamefont {Chen}}]{Huang2015}%
  \BibitemOpen
  \bibfield  {author} {\bibinfo {author} {\bibfnamefont {X.}~\bibnamefont
  {Huang}}, \bibinfo {author} {\bibfnamefont {L.}~\bibnamefont {Zhao}},
  \bibinfo {author} {\bibfnamefont {Y.}~\bibnamefont {Long}}, \bibinfo {author}
  {\bibfnamefont {P.}~\bibnamefont {Wang}}, \bibinfo {author} {\bibfnamefont
  {D.}~\bibnamefont {Chen}}, \bibinfo {author} {\bibfnamefont {Z.}~\bibnamefont
  {Yang}}, \bibinfo {author} {\bibfnamefont {H.}~\bibnamefont {Liang}},
  \bibinfo {author} {\bibfnamefont {M.}~\bibnamefont {Xue}}, \bibinfo {author}
  {\bibfnamefont {H.}~\bibnamefont {Weng}}, \bibinfo {author} {\bibfnamefont
  {Z.}~\bibnamefont {Fang}}, \bibinfo {author} {\bibfnamefont {X.}~\bibnamefont
  {Dai}},\ and\ \bibinfo {author} {\bibfnamefont {G.}~\bibnamefont {Chen}},\
  }\bibfield  {title} {\bibinfo {title} {Observation of the
  chiral-anomaly-induced negative magnetoresistance in 3d weyl semimetal
  taas},\ }\href {https://doi.org/10.1103/PhysRevX.5.031023} {\bibfield
  {journal} {\bibinfo  {journal} {Phys. Rev. X}\ }\textbf {\bibinfo {volume}
  {5}},\ \bibinfo {pages} {031023} (\bibinfo {year} {2015})}\BibitemShut
  {NoStop}%
\bibitem [{\citenamefont {Li}\ \emph {et~al.}(2018{\natexlab{a}})\citenamefont
  {Li}, \citenamefont {Wang}, \citenamefont {He}, \citenamefont {Wang},\ and\
  \citenamefont {Shen}}]{Li2018}%
  \BibitemOpen
  \bibfield  {author} {\bibinfo {author} {\bibfnamefont {H.}~\bibnamefont
  {Li}}, \bibinfo {author} {\bibfnamefont {H.-W.}\ \bibnamefont {Wang}},
  \bibinfo {author} {\bibfnamefont {H.}~\bibnamefont {He}}, \bibinfo {author}
  {\bibfnamefont {J.}~\bibnamefont {Wang}},\ and\ \bibinfo {author}
  {\bibfnamefont {S.-Q.}\ \bibnamefont {Shen}},\ }\bibfield  {title} {\bibinfo
  {title} {Giant anisotropic magnetoresistance and planar hall effect in the
  dirac semimetal ${\mathrm{cd}}_{3}{\mathrm{as}}_{2}$},\ }\href
  {https://doi.org/10.1103/PhysRevB.97.201110} {\bibfield  {journal} {\bibinfo
  {journal} {Phys. Rev. B}\ }\textbf {\bibinfo {volume} {97}},\ \bibinfo
  {pages} {201110} (\bibinfo {year} {2018}{\natexlab{a}})}\BibitemShut
  {NoStop}%
\bibitem [{\citenamefont {Burkov}(2017)}]{Burkov2017}%
  \BibitemOpen
  \bibfield  {author} {\bibinfo {author} {\bibfnamefont {A.~A.}\ \bibnamefont
  {Burkov}},\ }\bibfield  {title} {\bibinfo {title} {Giant planar hall effect
  in topological metals},\ }\href {https://doi.org/10.1103/PhysRevB.96.041110}
  {\bibfield  {journal} {\bibinfo  {journal} {Phys. Rev. B}\ }\textbf {\bibinfo
  {volume} {96}},\ \bibinfo {pages} {041110} (\bibinfo {year}
  {2017})}\BibitemShut {NoStop}%
\bibitem [{\citenamefont {Nandy}\ \emph {et~al.}(2017)\citenamefont {Nandy},
  \citenamefont {Sharma}, \citenamefont {Taraphder},\ and\ \citenamefont
  {Tewari}}]{Nandy2017}%
  \BibitemOpen
  \bibfield  {author} {\bibinfo {author} {\bibfnamefont {S.}~\bibnamefont
  {Nandy}}, \bibinfo {author} {\bibfnamefont {G.}~\bibnamefont {Sharma}},
  \bibinfo {author} {\bibfnamefont {A.}~\bibnamefont {Taraphder}},\ and\
  \bibinfo {author} {\bibfnamefont {S.}~\bibnamefont {Tewari}},\ }\bibfield
  {title} {\bibinfo {title} {Chiral anomaly as the origin of the planar hall
  effect in weyl semimetals},\ }\href
  {https://doi.org/10.1103/PhysRevLett.119.176804} {\bibfield  {journal}
  {\bibinfo  {journal} {Phys. Rev. Lett.}\ }\textbf {\bibinfo {volume} {119}},\
  \bibinfo {pages} {176804} (\bibinfo {year} {2017})}\BibitemShut {NoStop}%
\bibitem [{\citenamefont {Wu}\ \emph {et~al.}(2018)\citenamefont {Wu},
  \citenamefont {Zheng}, \citenamefont {Chu}, \citenamefont {Liu},
  \citenamefont {Gao}, \citenamefont {Zhang}, \citenamefont {Lu}, \citenamefont
  {Han}, \citenamefont {Zhou}, \citenamefont {Ning},\ and\ \citenamefont
  {Tian}}]{Wu2018}%
  \BibitemOpen
  \bibfield  {author} {\bibinfo {author} {\bibfnamefont {M.}~\bibnamefont
  {Wu}}, \bibinfo {author} {\bibfnamefont {G.}~\bibnamefont {Zheng}}, \bibinfo
  {author} {\bibfnamefont {W.}~\bibnamefont {Chu}}, \bibinfo {author}
  {\bibfnamefont {Y.}~\bibnamefont {Liu}}, \bibinfo {author} {\bibfnamefont
  {W.}~\bibnamefont {Gao}}, \bibinfo {author} {\bibfnamefont {H.}~\bibnamefont
  {Zhang}}, \bibinfo {author} {\bibfnamefont {J.}~\bibnamefont {Lu}}, \bibinfo
  {author} {\bibfnamefont {Y.}~\bibnamefont {Han}}, \bibinfo {author}
  {\bibfnamefont {J.}~\bibnamefont {Zhou}}, \bibinfo {author} {\bibfnamefont
  {W.}~\bibnamefont {Ning}},\ and\ \bibinfo {author} {\bibfnamefont
  {M.}~\bibnamefont {Tian}},\ }\bibfield  {title} {\bibinfo {title} {Probing
  the chiral anomaly by planar hall effect in dirac semimetal
  ${\mathrm{cd}}_{3}{\mathrm{as}}_{2}$ nanoplates},\ }\href
  {https://doi.org/10.1103/PhysRevB.98.161110} {\bibfield  {journal} {\bibinfo
  {journal} {Phys. Rev. B}\ }\textbf {\bibinfo {volume} {98}},\ \bibinfo
  {pages} {161110} (\bibinfo {year} {2018})}\BibitemShut {NoStop}%
\bibitem [{\citenamefont {Liang}\ \emph {et~al.}(2018)\citenamefont {Liang},
  \citenamefont {Lin}, \citenamefont {Kushwaha}, \citenamefont {Xing},
  \citenamefont {Ni}, \citenamefont {Cava},\ and\ \citenamefont
  {Ong}}]{Liang2018}%
  \BibitemOpen
  \bibfield  {author} {\bibinfo {author} {\bibfnamefont {S.}~\bibnamefont
  {Liang}}, \bibinfo {author} {\bibfnamefont {J.}~\bibnamefont {Lin}}, \bibinfo
  {author} {\bibfnamefont {S.}~\bibnamefont {Kushwaha}}, \bibinfo {author}
  {\bibfnamefont {J.}~\bibnamefont {Xing}}, \bibinfo {author} {\bibfnamefont
  {N.}~\bibnamefont {Ni}}, \bibinfo {author} {\bibfnamefont {R.~J.}\
  \bibnamefont {Cava}},\ and\ \bibinfo {author} {\bibfnamefont {N.~P.}\
  \bibnamefont {Ong}},\ }\bibfield  {title} {\bibinfo {title} {Experimental
  tests of the chiral anomaly magnetoresistance in the dirac-weyl semimetals
  ${\mathrm{na}}_{3}\mathrm{Bi}$ and gdptbi},\ }\href
  {https://doi.org/10.1103/PhysRevX.8.031002} {\bibfield  {journal} {\bibinfo
  {journal} {Phys. Rev. X}\ }\textbf {\bibinfo {volume} {8}},\ \bibinfo {pages}
  {031002} (\bibinfo {year} {2018})}\BibitemShut {NoStop}%
\bibitem [{\citenamefont {Chen}\ \emph {et~al.}(2018)\citenamefont {Chen},
  \citenamefont {Luo}, \citenamefont {Yan}, \citenamefont {Sun}, \citenamefont
  {Lv}, \citenamefont {Lu}, \citenamefont {Xi}, \citenamefont {Tong},
  \citenamefont {Sheng}, \citenamefont {Zhu}, \citenamefont {Song},\ and\
  \citenamefont {Sun}}]{Chen2018}%
  \BibitemOpen
  \bibfield  {author} {\bibinfo {author} {\bibfnamefont {F.~C.}\ \bibnamefont
  {Chen}}, \bibinfo {author} {\bibfnamefont {X.}~\bibnamefont {Luo}}, \bibinfo
  {author} {\bibfnamefont {J.}~\bibnamefont {Yan}}, \bibinfo {author}
  {\bibfnamefont {Y.}~\bibnamefont {Sun}}, \bibinfo {author} {\bibfnamefont
  {H.~Y.}\ \bibnamefont {Lv}}, \bibinfo {author} {\bibfnamefont {W.~J.}\
  \bibnamefont {Lu}}, \bibinfo {author} {\bibfnamefont {C.~Y.}\ \bibnamefont
  {Xi}}, \bibinfo {author} {\bibfnamefont {P.}~\bibnamefont {Tong}}, \bibinfo
  {author} {\bibfnamefont {Z.~G.}\ \bibnamefont {Sheng}}, \bibinfo {author}
  {\bibfnamefont {X.~B.}\ \bibnamefont {Zhu}}, \bibinfo {author} {\bibfnamefont
  {W.~H.}\ \bibnamefont {Song}},\ and\ \bibinfo {author} {\bibfnamefont
  {Y.~P.}\ \bibnamefont {Sun}},\ }\bibfield  {title} {\bibinfo {title} {Planar
  hall effect in the type-ii weyl semimetal
  ${T}_{d}\text{\ensuremath{-}}\mathrm{MoT}{\mathrm{e}}_{2}$},\ }\href
  {https://doi.org/10.1103/PhysRevB.98.041114} {\bibfield  {journal} {\bibinfo
  {journal} {Phys. Rev. B}\ }\textbf {\bibinfo {volume} {98}},\ \bibinfo
  {pages} {041114} (\bibinfo {year} {2018})}\BibitemShut {NoStop}%
\bibitem [{\citenamefont {Li}\ \emph {et~al.}(2018{\natexlab{b}})\citenamefont
  {Li}, \citenamefont {Zhang}, \citenamefont {Zhang}, \citenamefont {Wen},\
  and\ \citenamefont {Zhang}}]{PLi2018}%
  \BibitemOpen
  \bibfield  {author} {\bibinfo {author} {\bibfnamefont {P.}~\bibnamefont
  {Li}}, \bibinfo {author} {\bibfnamefont {C.~H.}\ \bibnamefont {Zhang}},
  \bibinfo {author} {\bibfnamefont {J.~W.}\ \bibnamefont {Zhang}}, \bibinfo
  {author} {\bibfnamefont {Y.}~\bibnamefont {Wen}},\ and\ \bibinfo {author}
  {\bibfnamefont {X.~X.}\ \bibnamefont {Zhang}},\ }\bibfield  {title} {\bibinfo
  {title} {Giant planar hall effect in the dirac semimetal
  $\mathrm{ZrT}{\mathrm{e}}_{5\ensuremath{-}\ensuremath{\delta}}$},\ }\href
  {https://doi.org/10.1103/PhysRevB.98.121108} {\bibfield  {journal} {\bibinfo
  {journal} {Phys. Rev. B}\ }\textbf {\bibinfo {volume} {98}},\ \bibinfo
  {pages} {121108} (\bibinfo {year} {2018}{\natexlab{b}})}\BibitemShut
  {NoStop}%
\bibitem [{\citenamefont {Singha}\ \emph {et~al.}(2018)\citenamefont {Singha},
  \citenamefont {Roy}, \citenamefont {Pariari}, \citenamefont {Satpati},\ and\
  \citenamefont {Mandal}}]{Singha2018}%
  \BibitemOpen
  \bibfield  {author} {\bibinfo {author} {\bibfnamefont {R.}~\bibnamefont
  {Singha}}, \bibinfo {author} {\bibfnamefont {S.}~\bibnamefont {Roy}},
  \bibinfo {author} {\bibfnamefont {A.}~\bibnamefont {Pariari}}, \bibinfo
  {author} {\bibfnamefont {B.}~\bibnamefont {Satpati}},\ and\ \bibinfo {author}
  {\bibfnamefont {P.}~\bibnamefont {Mandal}},\ }\bibfield  {title} {\bibinfo
  {title} {Planar hall effect in the type-ii dirac semimetal
  ${\mathrm{val}}_{3}$},\ }\href {https://doi.org/10.1103/PhysRevB.98.081103}
  {\bibfield  {journal} {\bibinfo  {journal} {Phys. Rev. B}\ }\textbf {\bibinfo
  {volume} {98}},\ \bibinfo {pages} {081103} (\bibinfo {year}
  {2018})}\BibitemShut {NoStop}%
\bibitem [{\citenamefont {Kumar}\ \emph {et~al.}(2018)\citenamefont {Kumar},
  \citenamefont {Guin}, \citenamefont {Felser},\ and\ \citenamefont
  {Shekhar}}]{Kumar2018}%
  \BibitemOpen
  \bibfield  {author} {\bibinfo {author} {\bibfnamefont {N.}~\bibnamefont
  {Kumar}}, \bibinfo {author} {\bibfnamefont {S.~N.}\ \bibnamefont {Guin}},
  \bibinfo {author} {\bibfnamefont {C.}~\bibnamefont {Felser}},\ and\ \bibinfo
  {author} {\bibfnamefont {C.}~\bibnamefont {Shekhar}},\ }\bibfield  {title}
  {\bibinfo {title} {Planar hall effect in the weyl semimetal gdptbi},\ }\href
  {https://doi.org/10.1103/PhysRevB.98.041103} {\bibfield  {journal} {\bibinfo
  {journal} {Phys. Rev. B}\ }\textbf {\bibinfo {volume} {98}},\ \bibinfo
  {pages} {041103} (\bibinfo {year} {2018})}\BibitemShut {NoStop}%
\bibitem [{\citenamefont {Yang}\ \emph {et~al.}(2020)\citenamefont {Yang},
  \citenamefont {Chang},\ and\ \citenamefont {Parkin}}]{Yang2020}%
  \BibitemOpen
  \bibfield  {author} {\bibinfo {author} {\bibfnamefont {S.-Y.}\ \bibnamefont
  {Yang}}, \bibinfo {author} {\bibfnamefont {K.}~\bibnamefont {Chang}},\ and\
  \bibinfo {author} {\bibfnamefont {S.~S.~P.}\ \bibnamefont {Parkin}},\
  }\bibfield  {title} {\bibinfo {title} {Large planar hall effect in bismuth
  thin films},\ }\href {https://doi.org/10.1103/PhysRevResearch.2.022029}
  {\bibfield  {journal} {\bibinfo  {journal} {Phys. Rev. Research}\ }\textbf
  {\bibinfo {volume} {2}},\ \bibinfo {pages} {022029} (\bibinfo {year}
  {2020})}\BibitemShut {NoStop}%
\bibitem [{\citenamefont {Owada}\ \emph {et~al.}(2018)\citenamefont {Owada},
  \citenamefont {Awashima},\ and\ \citenamefont {Fuseya}}]{Owada2018}%
  \BibitemOpen
  \bibfield  {author} {\bibinfo {author} {\bibfnamefont {M.}~\bibnamefont
  {Owada}}, \bibinfo {author} {\bibfnamefont {Y.}~\bibnamefont {Awashima}},\
  and\ \bibinfo {author} {\bibfnamefont {Y.}~\bibnamefont {Fuseya}},\
  }\bibfield  {title} {\bibinfo {title} {Corrections to the magnetoresistance
  formula for semimetals with dirac electrons: the boltzmann equation approach
  validated by the kubo formula},\ }\href
  {https://doi.org/10.1088/1361-648x/aae03c} {\bibfield  {journal} {\bibinfo
  {journal} {Journal of Physics: Condensed Matter}\ }\textbf {\bibinfo {volume}
  {30}},\ \bibinfo {pages} {445601} (\bibinfo {year} {2018})}\BibitemShut
  {NoStop}%
\bibitem [{\citenamefont {Dresselhaus}(1971)}]{Dresselhaus1971}%
  \BibitemOpen
  \bibfield  {author} {\bibinfo {author} {\bibfnamefont {M.~S.}\ \bibnamefont
  {Dresselhaus}},\ }\bibfield  {title} {\bibinfo {title} {Electronic properties
  of the group v semimetals},\ }\href@noop {} {\bibfield  {journal} {\bibinfo
  {journal} {J. Phys. Chem. Solids}\ }\textbf {\bibinfo {volume} {32}},\
  \bibinfo {pages} {3} (\bibinfo {year} {1971})}\BibitemShut {NoStop}%
\bibitem [{\citenamefont {\'Edel'man}(1976)}]{Edelman1976}%
  \BibitemOpen
  \bibfield  {author} {\bibinfo {author} {\bibfnamefont {V.~S.}\ \bibnamefont
  {\'Edel'man}},\ }\bibfield  {title} {\bibinfo {title} {Electrons in
  bismuth},\ }\href@noop {} {\bibfield  {journal} {\bibinfo  {journal} {Adv.
  Phys.}\ }\textbf {\bibinfo {volume} {25}},\ \bibinfo {pages} {555} (\bibinfo
  {year} {1976})}\BibitemShut {NoStop}%
\bibitem [{\citenamefont {Issi}(1979)}]{Issi1979}%
  \BibitemOpen
  \bibfield  {author} {\bibinfo {author} {\bibfnamefont {J.~P.}\ \bibnamefont
  {Issi}},\ }\bibfield  {title} {\bibinfo {title} {Low temperature transport
  properties of the group v semimetals},\ }\href@noop {} {\bibfield  {journal}
  {\bibinfo  {journal} {Aust. J. Phys.}\ }\textbf {\bibinfo {volume} {32}},\
  \bibinfo {pages} {585} (\bibinfo {year} {1979})}\BibitemShut {NoStop}%
\bibitem [{\citenamefont {Fuseya}\ \emph
  {et~al.}(2015{\natexlab{a}})\citenamefont {Fuseya}, \citenamefont {Ogata},\
  and\ \citenamefont {Fukuyama}}]{Fuseya2015}%
  \BibitemOpen
  \bibfield  {author} {\bibinfo {author} {\bibfnamefont {Y.}~\bibnamefont
  {Fuseya}}, \bibinfo {author} {\bibfnamefont {M.}~\bibnamefont {Ogata}},\ and\
  \bibinfo {author} {\bibfnamefont {H.}~\bibnamefont {Fukuyama}},\ }\bibfield
  {title} {\bibinfo {title} {Transport properties and diamagnetism of dirac
  electrons in bismuth},\ }\href@noop {} {\bibfield  {journal} {\bibinfo
  {journal} {J. Phys. Soc. Jpn.}\ }\textbf {\bibinfo {volume} {84}},\ \bibinfo
  {pages} {012001} (\bibinfo {year} {2015}{\natexlab{a}})}\BibitemShut
  {NoStop}%
\bibitem [{\citenamefont {Zhu}\ \emph {et~al.}(2018)\citenamefont {Zhu},
  \citenamefont {Fauqu{\'e}}, \citenamefont {Behnia},\ and\ \citenamefont
  {Fuseya}}]{Zhu2018}%
  \BibitemOpen
  \bibfield  {author} {\bibinfo {author} {\bibfnamefont {Z.}~\bibnamefont
  {Zhu}}, \bibinfo {author} {\bibfnamefont {B.}~\bibnamefont {Fauqu{\'e}}},
  \bibinfo {author} {\bibfnamefont {K.}~\bibnamefont {Behnia}},\ and\ \bibinfo
  {author} {\bibfnamefont {Y.}~\bibnamefont {Fuseya}},\ }\bibfield  {title}
  {\bibinfo {title} {Magnetoresistance and valley degree of freedom in bulk
  bismuth},\ }\href {http://stacks.iop.org/0953-8984/30/i=31/a=313001}
  {\bibfield  {journal} {\bibinfo  {journal} {J. Phys.: Condens. Matter}\
  }\textbf {\bibinfo {volume} {30}},\ \bibinfo {pages} {313001} (\bibinfo
  {year} {2018})}\BibitemShut {NoStop}%
\bibitem [{\citenamefont {Collaudin}\ \emph {et~al.}(2015)\citenamefont
  {Collaudin}, \citenamefont {Fauqu\'e}, \citenamefont {Fuseya}, \citenamefont
  {Kang},\ and\ \citenamefont {Behnia}}]{Collaudin2015}%
  \BibitemOpen
  \bibfield  {author} {\bibinfo {author} {\bibfnamefont {A.}~\bibnamefont
  {Collaudin}}, \bibinfo {author} {\bibfnamefont {B.}~\bibnamefont {Fauqu\'e}},
  \bibinfo {author} {\bibfnamefont {Y.}~\bibnamefont {Fuseya}}, \bibinfo
  {author} {\bibfnamefont {W.}~\bibnamefont {Kang}},\ and\ \bibinfo {author}
  {\bibfnamefont {K.}~\bibnamefont {Behnia}},\ }\bibfield  {title} {\bibinfo
  {title} {Angle dependence of the orbital magnetoresistance in bismuth},\
  }\href {https://doi.org/10.1103/PhysRevX.5.021022} {\bibfield  {journal}
  {\bibinfo  {journal} {Phys. Rev. X}\ }\textbf {\bibinfo {volume} {5}},\
  \bibinfo {pages} {021022} (\bibinfo {year} {2015})}\BibitemShut {NoStop}%
\bibitem [{\citenamefont {Zhu}\ \emph {et~al.}(2017)\citenamefont {Zhu},
  \citenamefont {Wang}, \citenamefont {Zuo}, \citenamefont {Fauqu{\'e}},
  \citenamefont {McDonald}, \citenamefont {Fuseya},\ and\ \citenamefont
  {Behnia}}]{Zhu2017}%
  \BibitemOpen
  \bibfield  {author} {\bibinfo {author} {\bibfnamefont {Z.}~\bibnamefont
  {Zhu}}, \bibinfo {author} {\bibfnamefont {J.}~\bibnamefont {Wang}}, \bibinfo
  {author} {\bibfnamefont {H.}~\bibnamefont {Zuo}}, \bibinfo {author}
  {\bibfnamefont {B.}~\bibnamefont {Fauqu{\'e}}}, \bibinfo {author}
  {\bibfnamefont {R.~D.}\ \bibnamefont {McDonald}}, \bibinfo {author}
  {\bibfnamefont {Y.}~\bibnamefont {Fuseya}},\ and\ \bibinfo {author}
  {\bibfnamefont {K.}~\bibnamefont {Behnia}},\ }\bibfield  {title} {\bibinfo
  {title} {Emptying dirac valleys in bismuth using high magnetic fields},\
  }\href {http://dx.doi.org/10.1038/ncomms15297} {\bibfield  {journal}
  {\bibinfo  {journal} {Nature Communications}\ }\textbf {\bibinfo {volume}
  {8}},\ \bibinfo {pages} {15297} (\bibinfo {year} {2017})}\BibitemShut
  {NoStop}%
\bibitem [{\citenamefont {Zhu}\ \emph {et~al.}(2012)\citenamefont {Zhu},
  \citenamefont {Collaudin}, \citenamefont {Fauqu{\'e}}, \citenamefont {Kang},\
  and\ \citenamefont {Behnia}}]{Zhu2012b}%
  \BibitemOpen
  \bibfield  {author} {\bibinfo {author} {\bibfnamefont {Z.}~\bibnamefont
  {Zhu}}, \bibinfo {author} {\bibfnamefont {A.}~\bibnamefont {Collaudin}},
  \bibinfo {author} {\bibfnamefont {B.}~\bibnamefont {Fauqu{\'e}}}, \bibinfo
  {author} {\bibfnamefont {W.}~\bibnamefont {Kang}},\ and\ \bibinfo {author}
  {\bibfnamefont {K.}~\bibnamefont {Behnia}},\ }\bibfield  {title} {\bibinfo
  {title} {Field-induced polarization of dirac valleys in bismuth},\ }\href
  {https://doi.org/10.1038/nphys2111} {\bibfield  {journal} {\bibinfo
  {journal} {Nature Physics}\ }\textbf {\bibinfo {volume} {8}},\ \bibinfo
  {pages} {89} (\bibinfo {year} {2012})}\BibitemShut {NoStop}%
\bibitem [{\citenamefont {Hartman}(1969)}]{Hartman1969}%
  \BibitemOpen
  \bibfield  {author} {\bibinfo {author} {\bibfnamefont {R.}~\bibnamefont
  {Hartman}},\ }\bibfield  {title} {\bibinfo {title} {Temperature dependence of
  the low-field galvanomagnetic coefficients of bismuth},\ }\href
  {https://doi.org/10.1103/PhysRev.181.1070} {\bibfield  {journal} {\bibinfo
  {journal} {Phys. Rev.}\ }\textbf {\bibinfo {volume} {181}},\ \bibinfo {pages}
  {1070} (\bibinfo {year} {1969})}\BibitemShut {NoStop}%
\bibitem [{\citenamefont {Smith}\ \emph {et~al.}(1964)\citenamefont {Smith},
  \citenamefont {Baraff},\ and\ \citenamefont {Rowell}}]{Smith1964}%
  \BibitemOpen
  \bibfield  {author} {\bibinfo {author} {\bibfnamefont {G.~E.}\ \bibnamefont
  {Smith}}, \bibinfo {author} {\bibfnamefont {G.~A.}\ \bibnamefont {Baraff}},\
  and\ \bibinfo {author} {\bibfnamefont {J.~M.}\ \bibnamefont {Rowell}},\
  }\bibfield  {title} {\bibinfo {title} {Effective $g$ factor of electrons and
  holes in bismuth},\ }\href@noop {} {\bibfield  {journal} {\bibinfo  {journal}
  {Phys. Rev.}\ }\textbf {\bibinfo {volume} {135}},\ \bibinfo {pages} {A1118}
  (\bibinfo {year} {1964})}\BibitemShut {NoStop}%
\bibitem [{\citenamefont {Hiruma}\ and\ \citenamefont
  {Miura}(1983)}]{Hiruma1983}%
  \BibitemOpen
  \bibfield  {author} {\bibinfo {author} {\bibfnamefont {K.}~\bibnamefont
  {Hiruma}}\ and\ \bibinfo {author} {\bibfnamefont {N.}~\bibnamefont {Miura}},\
  }\bibfield  {title} {\bibinfo {title} {Magnetoresistance study of bi and
  bi--sb alloys in high magnetic fields. ii. landau levels and
  semimetal-semiconductor transition},\ }\href
  {https://doi.org/10.1143/JPSJ.52.2118} {\bibfield  {journal} {\bibinfo
  {journal} {Journal of the Physical Society of Japan}\ }\textbf {\bibinfo
  {volume} {52}},\ \bibinfo {pages} {2118} (\bibinfo {year}
  {1983})}\BibitemShut {NoStop}%
\bibitem [{\citenamefont {Zhu}\ \emph {et~al.}(2011)\citenamefont {Zhu},
  \citenamefont {Fauqu\'e}, \citenamefont {Fuseya},\ and\ \citenamefont
  {Behnia}}]{Zhu2011}%
  \BibitemOpen
  \bibfield  {author} {\bibinfo {author} {\bibfnamefont {Z.}~\bibnamefont
  {Zhu}}, \bibinfo {author} {\bibfnamefont {B.}~\bibnamefont {Fauqu\'e}},
  \bibinfo {author} {\bibfnamefont {Y.}~\bibnamefont {Fuseya}},\ and\ \bibinfo
  {author} {\bibfnamefont {K.}~\bibnamefont {Behnia}},\ }\bibfield  {title}
  {\bibinfo {title} {Angle resolved landau spectrum of electrons and holes in
  bismuth},\ }\href@noop {} {\bibfield  {journal} {\bibinfo  {journal} {Phys.
  Rev. B}\ }\textbf {\bibinfo {volume} {84}},\ \bibinfo {pages} {115137}
  (\bibinfo {year} {2011})}\BibitemShut {NoStop}%
\bibitem [{\citenamefont {Aubrey}(1971)}]{Aubrey1971}%
  \BibitemOpen
  \bibfield  {author} {\bibinfo {author} {\bibfnamefont {J.~E.}\ \bibnamefont
  {Aubrey}},\ }\bibfield  {title} {\bibinfo {title} {Magnetoconductivity tensor
  for semimetals},\ }\href {http://stacks.iop.org/0305-4608/1/i=4/a=321}
  {\bibfield  {journal} {\bibinfo  {journal} {Journal of Physics F: Metal
  Physics}\ }\textbf {\bibinfo {volume} {1}},\ \bibinfo {pages} {493} (\bibinfo
  {year} {1971})}\BibitemShut {NoStop}%
\bibitem [{\citenamefont {Roth}(1992)}]{Roth1992}%
  \BibitemOpen
  \bibfield  {author} {\bibinfo {author} {\bibfnamefont {L.~M.}\ \bibnamefont
  {Roth}},\ }\bibinfo {title} {Dynamics and classical transport of carriers in
  semiconductors}\ (\bibinfo  {publisher} {North Holland},\ \bibinfo {year}
  {1992})\ Chap.~\bibinfo {chapter} {10}, p.\ \bibinfo {pages}
  {489}\BibitemShut {NoStop}%
\bibitem [{\citenamefont {Askerov}(1994)}]{Askerov1994}%
  \BibitemOpen
  \bibfield  {author} {\bibinfo {author} {\bibfnamefont {B.~M.}\ \bibnamefont
  {Askerov}},\ }\href {https://doi.org/10.1142/1926} {\emph {\bibinfo {title}
  {Electron Transport Phenomena in Semiconductors}}}\ (\bibinfo  {publisher}
  {WORLD SCIENTIFIC},\ \bibinfo {year} {1994})\BibitemShut {NoStop}%
\bibitem [{\citenamefont {Mackey}\ and\ \citenamefont
  {Sybert}(1969)}]{Mackey1969}%
  \BibitemOpen
  \bibfield  {author} {\bibinfo {author} {\bibfnamefont {H.~J.}\ \bibnamefont
  {Mackey}}\ and\ \bibinfo {author} {\bibfnamefont {J.~R.}\ \bibnamefont
  {Sybert}},\ }\bibfield  {title} {\bibinfo {title} {Magnetoconductivity of a
  fermi ellipsoid with anisotropic relaxation time},\ }\href
  {https://doi.org/10.1103/PhysRev.180.678} {\bibfield  {journal} {\bibinfo
  {journal} {Phys. Rev.}\ }\textbf {\bibinfo {volume} {180}},\ \bibinfo {pages}
  {678} (\bibinfo {year} {1969})}\BibitemShut {NoStop}%
\bibitem [{\citenamefont {Wolff}(1964)}]{Wolff1964}%
  \BibitemOpen
  \bibfield  {author} {\bibinfo {author} {\bibfnamefont {P.~A.}\ \bibnamefont
  {Wolff}},\ }\href@noop {} {\bibfield  {journal} {\bibinfo  {journal} {J.
  Phys. Chem. Solids}\ }\textbf {\bibinfo {volume} {25}},\ \bibinfo {pages}
  {1057} (\bibinfo {year} {1964})}\BibitemShut {NoStop}%
\bibitem [{\citenamefont {Baraff}(1965)}]{Baraff1965}%
  \BibitemOpen
  \bibfield  {author} {\bibinfo {author} {\bibfnamefont {G.~A.}\ \bibnamefont
  {Baraff}},\ }\bibfield  {title} {\bibinfo {title} {Magnetic energy levels in
  the bismuth conduction band},\ }\href@noop {} {\bibfield  {journal} {\bibinfo
   {journal} {Phys. Rev.}\ }\textbf {\bibinfo {volume} {137}},\ \bibinfo
  {pages} {A842} (\bibinfo {year} {1965})}\BibitemShut {NoStop}%
\bibitem [{\citenamefont {Fuseya}\ \emph
  {et~al.}(2015{\natexlab{b}})\citenamefont {Fuseya}, \citenamefont {Zhu},
  \citenamefont {Fauqu\'e}, \citenamefont {Kang}, \citenamefont {Lenoir},\ and\
  \citenamefont {Behnia}}]{Fuseya2015b}%
  \BibitemOpen
  \bibfield  {author} {\bibinfo {author} {\bibfnamefont {Y.}~\bibnamefont
  {Fuseya}}, \bibinfo {author} {\bibfnamefont {Z.}~\bibnamefont {Zhu}},
  \bibinfo {author} {\bibfnamefont {B.}~\bibnamefont {Fauqu\'e}}, \bibinfo
  {author} {\bibfnamefont {W.}~\bibnamefont {Kang}}, \bibinfo {author}
  {\bibfnamefont {B.}~\bibnamefont {Lenoir}},\ and\ \bibinfo {author}
  {\bibfnamefont {K.}~\bibnamefont {Behnia}},\ }\bibfield  {title} {\bibinfo
  {title} {Origin of the large anisotropic $g$ factor of holes in bismuth},\
  }\href {https://doi.org/10.1103/PhysRevLett.115.216401} {\bibfield  {journal}
  {\bibinfo  {journal} {Phys. Rev. Lett.}\ }\textbf {\bibinfo {volume} {115}},\
  \bibinfo {pages} {216401} (\bibinfo {year} {2015}{\natexlab{b}})}\BibitemShut
  {NoStop}%
\bibitem [{\citenamefont {Izaki}\ and\ \citenamefont
  {Fuseya}(2019)}]{Izaki2019}%
  \BibitemOpen
  \bibfield  {author} {\bibinfo {author} {\bibfnamefont {Y.}~\bibnamefont
  {Izaki}}\ and\ \bibinfo {author} {\bibfnamefont {Y.}~\bibnamefont {Fuseya}},\
  }\bibfield  {title} {\bibinfo {title} {Nonperturbative matrix mechanics
  approach to spin-split landau levels and the $g$ factor in spin-orbit coupled
  solids},\ }\href {https://doi.org/10.1103/PhysRevLett.123.156403} {\bibfield
  {journal} {\bibinfo  {journal} {Phys. Rev. Lett.}\ }\textbf {\bibinfo
  {volume} {123}},\ \bibinfo {pages} {156403} (\bibinfo {year}
  {2019})}\BibitemShut {NoStop}%
\bibitem [{See()}]{SeeSM}%
  \BibitemOpen
  \href@noop {} {}\bibinfo {note} {See Supplemental Material [url] for
  details.}\BibitemShut {Stop}%
\bibitem [{\citenamefont {Song}\ \emph {et~al.}(2015)\citenamefont {Song},
  \citenamefont {Refael},\ and\ \citenamefont {Lee}}]{Song2015}%
  \BibitemOpen
  \bibfield  {author} {\bibinfo {author} {\bibfnamefont {J.~C.~W.}\
  \bibnamefont {Song}}, \bibinfo {author} {\bibfnamefont {G.}~\bibnamefont
  {Refael}},\ and\ \bibinfo {author} {\bibfnamefont {P.~A.}\ \bibnamefont
  {Lee}},\ }\bibfield  {title} {\bibinfo {title} {Linear magnetoresistance in
  metals: Guiding center diffusion in a smooth random potential},\ }\href
  {https://doi.org/10.1103/PhysRevB.92.180204} {\bibfield  {journal} {\bibinfo
  {journal} {Phys. Rev. B}\ }\textbf {\bibinfo {volume} {92}},\ \bibinfo
  {pages} {180204} (\bibinfo {year} {2015})}\BibitemShut {NoStop}%
\bibitem [{\citenamefont {Abrikosov}(1969)}]{Abrikosov1969}%
  \BibitemOpen
  \bibfield  {author} {\bibinfo {author} {\bibfnamefont {A.~A.}\ \bibnamefont
  {Abrikosov}},\ }\bibfield  {title} {\bibinfo {title} {Galvanomagnetic
  phenomena in metals in the quantum limit},\ }\href@noop {} {\bibfield
  {journal} {\bibinfo  {journal} {Sov. Phys. JETP}\ }\textbf {\bibinfo {volume}
  {29}},\ \bibinfo {pages} {746} (\bibinfo {year} {1969})}\BibitemShut
  {NoStop}%
\bibitem [{\citenamefont {Mahan}(1983)}]{Mahan1983}%
  \BibitemOpen
  \bibfield  {author} {\bibinfo {author} {\bibfnamefont {G.~D.}\ \bibnamefont
  {Mahan}},\ }\bibfield  {title} {\bibinfo {title} {Linear magnetoresistance at
  high magnetic field},\ }\href {https://doi.org/10.1088/0305-4608/13/12/004}
  {\bibfield  {journal} {\bibinfo  {journal} {Journal of Physics F: Metal
  Physics}\ }\textbf {\bibinfo {volume} {13}},\ \bibinfo {pages} {L257}
  (\bibinfo {year} {1983})}\BibitemShut {NoStop}%
\end{thebibliography}%

\end{document}